 \documentclass[a4paper,10pt]{article}
\usepackage{graphicx,rotating,hyperref,slashed,amsmath,xcolor,amssymb,amsfonts,colortbl,cite}
\pdfoutput=1  
\makeatletter
\hypersetup{colorlinks,bookmarksopen,bookmarksnumbered,
linkcolor=blus,pdfstartview=FitH,urlcolor=rossos,citecolor=verde}
\allowdisplaybreaks

\def\lsim{\mathrel{\rlap{\lower3pt\hbox{\hskip0pt$\sim$}}
   \raise1pt\hbox{$<$}}}         
\def\gsim{\mathrel{\rlap{\lower4pt\hbox{\hskip1pt$\sim$}}
   \raise1pt\hbox{$>$}}}         

 \newcommand{\sfootnote}[1]{}
\definecolor{bluc}{cmyk}{1,1,0,0.1}
\definecolor{rossoCP3}{cmyk}{0,.88,.77,.40}
\definecolor{rosso}{cmyk}{0,1,1,0.4}
\definecolor{rossos}{cmyk}{0,1,1,0.55}
\definecolor{rossoc}{cmyk}{0,1,1,0.2}
\definecolor{verdes}{cmyk}{0.92,0,0.59,0.4}

\hypersetup{colorlinks, bookmarksopen, bookmarksnumbered,
citecolor=verdes, linkcolor=bluc, pdfstartview=FitH, urlcolor=rossos}

\newcommand{\mio}[1]{}

\definecolor{Gray}{gray}{0.95}

\usepackage{multicol}
\usepackage{color}
\definecolor{rosso}{cmyk}{0,1,1,0.4}
\definecolor{rossos}{cmyk}{0,1,1,0.55}
\definecolor{rossoc}{cmyk}{0,1,1,0.2}
\definecolor{blu}{cmyk}{1,1,0,0.3}
\definecolor{blus}{cmyk}{1,1,0,0.6}
\definecolor{bluc}{cmyk}{1,1,0,0.1}
\definecolor{verde}{cmyk}{0.92,0,0.59,0.25}
\definecolor{verdec}{cmyk}{0.92,0,0.59,0.15}
\definecolor{verdes}{cmyk}{0.92,0,0.59,0.4}

\def\circa#1{\,\raise.3ex\hbox{$#1$\kern-.75em\lower1ex\hbox{$\sim$}}\,}

\newcommand{\beq}{\begin{equation}}
\newcommand{\eeq}{\end{equation}}

\newcommand{\bea}{\begin{eqnarray}}
\newcommand{\eea}{\end{eqnarray}}
\newcommand{\be}{\begin{equation}}
\newcommand{\ee}{\end{equation}}
\newfam\rsfsfam
\def\mathscr#1{{\fam\rsfsfam\relax#1}}

\newcommand{\ba}{\begin{eqnarray}}
\newcommand{\ea}{\end{eqnarray}}

\def\circa#1{\,\raise.3ex\hbox{$#1$\kern-.75em\lower1ex\hbox{$\sim$}}\,}
\makeatletter

\def\hhref#1{\href{http://arxiv.org/abs/#1}{arXiv:#1}} 

\newcommand{\doi}[1]{\href{http://dx.doi.org/#1}{[doi]}}

\def\hhref#1{\href{http://arxiv.org/abs/#1}{arXiv:#1}} 
 
\def\art{\@ifnextchar[{\eart}{\oart}}
\def\eart[#1]#2#3#4#5#6{{\rm #2}, {\em #3 \bf #4} {\rm (#6) #5} ({\em #1})}

\def\article{\@ifnextchar[{\earticle}{\oarticle}}
\def\oarticle#1#2#3#4#5#6{{\rm #1}, {\em ``#6''}, {\rm #2 #3 (#5) #4}}
\def\earticle[#1]#2#3#4#5#6#7{{\rm #2}, {\em ``#7''}, {\rm #3 #4 (#6) #5}  [\hhref{#1}]}
\def\hepart[#1]#2{{\rm #2, \em#1}}
\def\heparticle[#1]#2#3{#2, {\em ``#3''} [\hhref{#1}]}

\newcounter{alphaequation}[equation]
\def\thealphaequation{\theequation\hbox to
0.6em{\hfil\alph{alphaequation}\hfil}}
\def\eqnsystem#1{
\def\@eqnnum{{\rm (\thealphaequation)}}
\def\@@eqncr{\let\@tempa\relax \ifcase\@eqcnt \def\@tempa{& & &} \or
  \def\@tempa{& &}\or \def\@tempa{&}\fi\@tempa
  \if@eqnsw\@eqnnum\refstepcounter{alphaequation}\fi
\global\@eqnswtrue\global\@eqcnt=0\cr}
\refstepcounter{equation} \let\@currentlabel\theequation \def\@tempb{#1}
\ifx\@tempb\empty\else\label{#1}\fi
\refstepcounter{alphaequation}
\let\@currentlabel\thealphaequation
\global\@eqnswtrue\global\@eqcnt=0 \tabskip\@centering\let\\=\@eqncr
$$\halign to \displaywidth\bgroup \@eqnsel\hskip\@centering
$\displaystyle\tabskip\z@{##}$&\global\@eqcnt\@ne
\hskip2\arraycolsep\hfil${##}$\hfil& \global\@eqcnt\tw@\hskip2\arraycolsep
$\displaystyle\tabskip\z@{##}$\hfil
\tabskip\@centering&\llap{##}\tabskip\z@\cr}
\def\endeqnsystem{\@@eqncr\egroup$$\global\@ignoretrue} \makeatother

\definecolor{fiorentina}{rgb}{.5,0,.5}

\begin{document}

\vspace{1truecm}
 
\begin{center}
\boldmath

{\textbf{\Large Stealth configurations  in vector-tensor theories of gravity
}}
\unboldmath

\unboldmath

\bigskip\bigskip

\vspace{0.1truecm}

{\bf Javier Chagoya,  Gianmassimo Tasinato }
 \\[8mm]
{\it Department of Physics, Swansea University, Swansea, SA2 8PP, U.K.}\\[1mm]

\vspace{1cm}

\thispagestyle{empty}
{\large\bf\color{blus} Abstract}
\begin{quote}

Studying the physics of compact objects in modified theories of gravity is important for understanding how future observations
can test alternatives to  General Relativity.  We consider  a subset of vector-tensor Galileon theories of gravity characterized
by new symmetries, which can prevent the propagation of the vector longitudinal polarization, even in absence of Abelian
gauge invariance.  We  investigate new 
    spherically symmetric and slowly rotating  solutions for these systems, including an arbitrary matter Lagrangian. We show that, under certain conditions, there  always exist stealth configurations whose geometry coincides with solutions of  Einstein gravity 
    coupled with the additional matter. Such solutions have   a non-trivial profile for the vector field, characterized
     by independent integration constants, which  extends to asymptotic infinity. 
      We interpret our findings in terms of the symmetries and features of the original vector-tensor action, and on the number of degrees of freedom that it propagates. 
      These results are  important  
    to eventually describe gravitationally bound configurations  
in modified theories of gravity, such as black holes and  neutron stars,   including   realistic matter fields forming or surrounding the object.

 \end{quote}
\thispagestyle{empty}
\end{center}

\setcounter{page}{1}
\setcounter{footnote}{0}



\section{Introduction}

The physics of black holes and other gravitationally bound objects can probe non-perturbative
aspects of gravitational theories, in regimes that deviate from weak-field approximations. This
is particularly important to explore modified gravity theories equipped with screening mechanisms, motivated
by the dark energy problem, 
which satisfy the strong constraints on deviations from General Relativity (GR) in the weak-field limit. See
\cite{Clifton:2011jh} for a review.  
The existence of non-trivial black hole solutions in scalar and vector-tensor theories of gravity is challenged
by powerful no-hair theorems, which can forbid the existence of solutions with scalar or vector hairs (see e.g. 
the nice review \cite{Bekenstein:1996pn}). 
 Ways out to these negative results exist, avoiding explicit or implicit assumptions at the base of the no-hair
 theorems.  For example, in shift symmetric Horndeski scalar-tensor theories \cite{Horndeski:1974wa}, a theorem by Hui and Nicolis
 generally forbids the existence of static black hole configurations with non-trivial scalar hair \cite{Hui:2012qt }. 
 Sotiriou and Zhou \cite{Sotiriou:2013qea, Sotiriou:2014pfa} found a way  to avoid one of the hypothesis of the theorem, by selecting particular, non-analytic forms 
 for the free functions characterizing  Horndeski theories. Many generalizations of these results have then been
 developed, see e.g. \cite{Herdeiro:2015waa,Sotiriou:2015pka,Volkov:2016ehx} for comprehensive reviews.

In this work, we investigate spherically symmetric and slowly rotating configurations in the vector-tensor Galileon
theories of gravity \cite{gripaios,tasinato,heisenberg} introduced to address the dark energy problem. In this case, Bekenstein classic no-hair theorems \cite{Bekenstein:1971hc} can be circumvented, leading to more
general families of black hole solutions, first found in \cite{Chagoya:2016aar} and then generalised in \cite{Minamitsuji:2016ydr}. Here we study 
theories with   non-analytic  choices of free functions characterizing them.   These cases
have not been much explored in the vector-tensor case, and have several interesting features that we point out
for the first time. In Section \ref{sec-sysc} we show that such choices of free functions are associated with
new symmetries, which can lead
 to systems  propagating less than the expected
number of degrees of freedom. The
  symmetries we identify allow us to single out specific vector-tensor Galileon systems, and we analyse the
properties of spherically symmetric and slowly rotating configurations in these systems. 
 In Section \ref{sec-solg}
we analyse a set-up where the  vector-tensor
action is accompanied by an Einstein-Hilbert contribution, as well as by an arbitrary, unspecified  matter Lagrangian,
with the only requirement that the latter does not directly couple with the vector field.
We prove that stealth solutions exist,  whose geometry  coincides exactly with the one found in Einstein gravity coupled with the matter
fields, but additionally  with a non-trivial profile for the vector with specific integration constants. We interpret these results
in terms of the properties of the original action, and the number of degrees of freedom that it propagates. We then show 
that by adding a standard Maxwell kinetic terms for the vector, one finds configurations that are small modifications of 
Reissner-Nordstr\"om black holes.  We sum-up in Section \ref{sec-concl}  with a discussion of possible further
developments.

\section{System under consideration} \label{sec-sysc}
This  Section introduces  the special class  of vector-tensor theories we examine, and find new 
  symmetries which can lead to systems that do not propagate the vector longitudinal mode.

\subsection{Special vector-tensor theories}
The vector-tensor theories dubbed vector Galileons  or generalized Proca \cite{gripaios,tasinato,heisenberg}  have been first introduced 
 as vector-tensor versions  of Galileon and Horndeski actions. Subsequent investigations started to
explore 
 their field theoretical \cite{fieldapp} and cosmological \cite{cosmoapp} ramifications.

The vector-tensor
Lagrangians we consider  are~\footnote{ We number the Lagrangians from $2$ to $5$ in analogy with the classification  of scalar-tensor  Horndeski theories: in the vector-tensor
   case, however, $L_{(1)}$ is absent. Also, in \cite{Allys:2015sht}, a higher order vector tensor Lagrangian $L_{(6)}$ was introduced: it does not play a role in our discussion, hence we do not consider it.  }:
\bea
L_{(2)}&=&\sqrt{-g}\,G_2, \label{lagr2}
\\
{ L}_{(3)} & = & \sqrt{-g}\,G_3\,\nabla_\mu A^\mu \label{lagr3}, \\
{ L}_{(4)} & = & \sqrt{-g}\,\left[
G_4 R +  G_{4,X}\left[(\nabla_\mu A^\mu)^2 - \nabla_\rho A_\sigma \nabla^\sigma A^\rho   \right] 
\right],
\\
 L_{(5)} & =& 
 \sqrt{-g}\,\left[
 G_5 G_{\mu\nu}\nabla^\mu A^\nu - \frac{1}{6} G_{5,X}\left[ (\nabla^\mu A_\mu)^3 - 3 \nabla^\mu A_\mu \nabla_\rho A_\sigma \nabla^\sigma A^\rho + 2 \nabla_\rho A_\sigma \nabla^\gamma A^\rho \nabla^\sigma A_\gamma \right]
\right], \label{lagr5}
\eea
where the $G_i$ are functions of $X$ only, and
\be 2 X = - A^\mu  A_\mu.
\ee
   We assume that $A_\mu$ has mass dimension one, as suggested by the standard canonical  normalization
   of  Maxwell  kinetic term. 
  If the vector is curl-free and has no transverse polarizations, 
 we can write $A_\mu\,=\,\partial_\mu \pi$ for a scalar $\pi$, and we obtain the {shift-symmetric} scalar-tensor 
 Horndeski theories.
    Since  Lagrangians \eqref{lagr2}-\eqref{lagr5} explicitly depend on the gauge potential $A_\mu$,   they
   break the Abelian $U(1)$ gauge symmetry.
   Around flat space,
   they
       generally propagate {\it five}
 degrees of freedom, which  can be decomposed into
  two tensors, two vector transverse modes, and one scalar (the vector longitudinal polarization).
        As for scalar Galileons, interactions have been selected
     such to avoid the propagation of a ghostly Ostrogradsky sixth mode 
  \cite{tasinato,heisenberg}. 
 
 There are special cases though, where new  interesting symmetries emerge.
    Consider the following specific choices
 of functions $G_i$ in eqs \eqref{lagr2}-\eqref{lagr5}:
 \bea
G_2&=&m^3\,\sqrt{X}\, , \label{g2h}
\\
G_3&=& m^2\,\ln{X}\, ,\\
G_4&=&m \,\sqrt{X}\, ,  \label{g4h}\\
G_5&=& \ln{X}\, , \label{g5h}
\eea
with  $m$ a parameter  with dimension of a mass. 
 Since these choices of functions $G_i$ are non analytic,  the
system 
  spontaneously breaks  Lorentz symmetry, and any consistent  vacuum requires $A_\mu \neq 0$.
 {This can be a relevant feature when applying}  these systems to cosmology, or when studying spherically symmetric
 configurations, as we are going to do in the second part of this work.  
 
 When the $G_i$ are chosen as in eqs \eqref{g2h}-\eqref{g5h}, up to overall constants and total derivatives, 
%
 the vector tensor Lagrangians  \eqref{lagr2}-\eqref{lagr5} can be re-expressed 
  as
 \be \label{LagCur}
 L_{(i)}\,=\,\sqrt{-g} \, m^{5-i}\,A_\mu\,J^\mu_{(i)} \, ,
 \ee
 with $i\,=\,2,..5$ and the vectors $J^\mu_{(i)}$
 given by 
 \begin{align} \label{cur-def2}
 J^\mu_{(2)}= & -{2 \, X^{-1/2}\,{A^\mu}}\, ,  \\
 J^\mu_{(3)} = &  -X^{-1} (A^\mu \nabla_\alpha A^\alpha - A^\alpha \nabla^\mu A_\alpha),  \\
 J^\mu_{(4)} = & X^{-1/2}\, G^{\mu\alpha} A_\alpha  -\frac{X^{-3/2}}{4} \epsilon^{\alpha\beta\sigma\delta}\epsilon_{\alpha}{}^{\mu\gamma\eta}  A_\beta \nabla_\sigma A_\gamma \nabla_\delta A _\eta 
 \,, \\
 J^\mu_{(5)} = &  
 \frac{1}{4 \,X} \epsilon^{\alpha\beta\sigma\delta}\epsilon^{\nu\rho\gamma\mu}  \left[  R_{\alpha\beta\rho\nu} - \frac{2}{3\,X}\nabla_\alpha A_\rho \nabla_\beta A_\nu   \right] A_\delta \nabla_\sigma A_\gamma  . 
 \end{align}
 Expressions \eqref{LagCur} for the Lagrangians are convenient since
 the vector equations of motion (eoms) are simply given 
   by  
   \be
   J^\mu_{(i)}\,=\,0 \hskip1cm {\text{(vector equations of motion).}}
   \ee 
  
  \bigskip
  The quantities $J^\mu_{(i)}$ are the vector analogues of the conserved currents in shift symmetric scalar-tensor Horndeski theories, see e.g. \cite{Hui:2012qt}. 
    The non analytic choice of functions  $G_i$  might  seem pathological or, at the very least, too specific for exploring 
  phenomenological consequences as properties of black hole solutions.
On the other hand,
  these vector tensor  theories have  important distinctive features,
     which further motivate their study. 
  They are candidates for the building blocks of
   `massless' set-ups, since can  
    propagate {\it less} than five degrees of freedom, even in absence of an  Abelian gauge symmetry. 
    We  support this statement  
     by means of two arguments, based on symmetry properties. In Section \ref{sym-dec} we analyse a decoupling limit,
     where a new global symmetry arises for all the theories above,  preventing
     a scalar mode to acquire a dynamics, and allowing the propagation of at most four degrees of freedom.  
     In  Section \ref{sym-out}, we show that some of the theories above enjoy a new symmetry even outside the decoupling
     limit, which  can forbid the propagation 
     of the vector longitudinal polarization around flat space.
%

 \subsection{Decoupling limit and associated symmetry}\label{sym-dec}
 
 We identify
 a regime where 
 scalar, vector, and tensor modes are decoupled, and
  the self-interactions of the 
   longitudinal 
  vector component can be isolated and clearly identified.
  Such decoupling limit allows us to make manifest a new
 symmetry characterizing the special Lagrangians that we consider,  which prevents the propagation
 of the scalar mode associated with the longitudinal vector degree of freedom.
   The study of similar decoupling limits
    have been essential in the past to identify the dynamics of degrees of freedom in modified gravity theories, such as dRGT massive
    gravity (see e.g. \cite{Hinterbichler:2011tt,deRham:2014zqa} for reviews). 
    In \cite{tasinato}, theories of vector Galileons were constructed by demanding that a decoupling limit 
  exists, where the action for the scalar  longitudinal vector mode obeys a Galileon symmetry. We 
  now discuss a     regime which leads to a different symmetry  for the  system
  we focus on.

  The reference action that we consider is 
  \be \label{effac1}
  S\,=\,\int d^4 x\, \left[\sqrt{-g}\,\left(
  M_{Pl}^2\,R-\frac14 F_{\mu\nu}^2 \right)+
   \sum_{i=2}^5 \lambda_i\,L_{(i)}  \right]\, ,
  \ee
 where the Lagrangian densities $L_{(i)}$ are  given in eqs \eqref{LagCur} (it is more convenient to  work with the Lagrangians as expressed in terms of currents).

  First of all, we restore an Abelian gauge symmetry in the system 
   by introducing  a 
   St\"uckelberg field. We make
    the following substitution 
 whenever we meet a vector potential in our action:
 \be
A_\mu \to A_\mu-\frac{1}{ m} \partial_\mu \pi \, ,
\ee
with $\pi$ a scalar  St\"uckelberg  field of mass dimension one, and for convenience  $m$ is the same parameter
appearing in the currents \eqref{cur-def2}. 
   After applying the St\"uckelberg trick,  action \eqref{effac1} is invariant under 
 the Abelian gauge symmetry 
 \be
 A_\mu\to A_\mu+\frac{1}{m}\partial_\mu \xi(x)\,\,,\hskip1cm\pi \to \pi+\xi(x)\,,
 \ee
 for an arbitrary function $\xi(x)$. 
 We now perturb the metric around Minkowski space:
 \be
 g_{\mu\nu}\,=\,\eta_{\mu\nu}+\frac{1}{M_{Pl}}\,h_{\mu\nu}\,,
 \ee
where $h_{\mu\nu}$ is the canonically normalised metric fluctuation. 

Decoupling  proceeds in two steps. First, we decouple gravity sending $M_{Pl}\to \infty$: all the dependence of 
the metric
 fluctuations  in
 the vector part of the action
  disappears. 
Then, we decouple the vector transverse modes
  from the vector longitudinal scalar mode. We  do so by considering the simultaneous limits
\bea
m&\to&0 \,,
\\
\lambda_i\,m^{4-i}&=&\beta_i \Lambda^{4-i} \hskip1cm {\text{remains finite,}} 
\eea
where $i\,=\,2,\dots 5$ is the index labelling the Lagrangians \eqref{LagCur},  $\Lambda$ a new energy scale, and
$\beta_i$ some new finite dimensionless parameters.
 Notice that,  when taking the limit $m\to0$, the constants $\lambda_i$ must
go to infinity (or to zero, depending on the value of $i$) in order to keep $\beta_i$ finite. 

We obtain the following decoupled action around Minkowski space:
\be \label{actdec}
S^{dec}\,=\,\int d^4 x\,\left[{\cal L}_{kin}(h_{\mu\nu})-\frac14 F_{\mu\nu}^2+
\sum_{i=2}^5 \beta_i\,L^{scal}_{(i)} \right]\,,
\ee
where ${\cal L}_{kin}(h_{\mu\nu})$ is the standard quadratic kinetic term for spin two tensor modes. Notice that the three sectors -- tensor, vector, scalar -- are decoupled as desired.
  $L^{scal}_{(i)} $ are the four scalar Lagrangians introduced in \cite{Chagoya:2016inc}:
\bea
L_{(2)}^{scal}&=& \Lambda^2 \,\sqrt{X_s} \label{lscal2},
\\
L_{(3)}^{scal}&=& \Lambda \left([\Pi]-\frac{1}{X_s} [\Phi] \right) \label{lscal3},
\\
L_{(4)}^{scal} & = &  \frac{1}{\sqrt{X_s}}\left( [\Pi]^2-[\Pi^2] +\frac{2}{X_s} \left([\Phi^2]- [\Phi] [\Pi] \right)\right)
 \label{lscal4}, 
 \\
 L_{(5)}^{scal} & = & \frac{1}{\Lambda\,X_s} \left(
[\Pi]^3+2 [\Pi^3]-3 [\Pi^2]  [\Pi] + \frac{3}{X_s}  \left(2 [\Pi] [\Phi^2]-2 [\Phi^3] - [\Phi] [\Pi]^2+
 [\Phi] [\Pi^2] \right)
\right)
 \label{lscal5}.
\eea
where $X_s\,=\,-\left(\partial_\mu \pi \partial^\mu \pi \right)/2$, 
$\Pi_{\mu\nu} = \nabla_\mu \nabla_\nu \pi$, $[\Pi^n] = {\rm{Tr}}\  \Pi^n$, and $[\Phi^n] = \partial \pi \cdot \Pi^n \cdot \partial\pi  $. 
Besides a constant shift symmetry, such Lagrangian densities are invariant -- up to total derivatives -- under the scalar symmetry
\be \label{scalsym1}
\delta \pi\,=\,\pi\,\omega^\mu \partial_\mu \pi\,,
\ee
with $\omega^\mu$ an arbitrary, constant four-vector.
This fact has been shown in \cite{Chagoya:2016inc}, to which we refer the reader for further details. Symmetry 
\eqref{scalsym1} is more manifest when embedding the system in a higher dimensional brane-world setting. It is   inherited by a bulk higher dimensional rotational symmetry,  spontaneously broken
by the presence of a brane, with $\pi$ playing the role of Goldstone boson (see also  \cite{Goon:2011qf} for the original papers developing these techniques).

\subsubsection*{No propagating scalar mode}

This symmetry is very constraining: in fact, it does not even allow for the propagation of standard scalar excitations. In our decoupling limit, we consider as reference metric  Minkowski space-time, and a vanishing profile for the vector transverse modes. The scalar equation of motion, and the non-analytic structure of  Lagrangians \eqref{lscal2}-\eqref{lscal5},
  require a time-like 
 non trivial profile for the scalar field, such that  $X_s\,>\,0$. 
 The scalar equation of motion associated with the decoupled action \eqref{actdec} contains
second derivatives of the scalar field, and any background configuration linear in coordinates is a solution:
we denote such background configuration
\be
\bar \pi(x)\,=\,c_\mu x^\mu\,,
\ee
for some arbitrary  time-like vector $c_\mu$.
We consider the case of  constant time-like vector $c_\mu$, and choose a frame where $c_\mu\,=\,(c_0,\,0,\,0,\,0)$, with $c_0$ constant,
so to preserve spatial isotropy.  
 This scalar background spontaneously  breaks Lorentz symmetry, and furthermore reduces 
the scalar symmetry \eqref{scalsym1} to 
\be \label{scalsym2}
\delta \pi\,=\,\pi\,\omega^a \partial_a \pi
\ee 
for an arbitrary three spatial vector $\omega^a$ (with $a\,=\,1,2,3$). This residual symmetry tells us much about the 
 (absence of) dynamics of scalar excitations. We denote perturbations around the background scalar profile as $\hat \pi$:
 \be
 \pi(x)\,=\,\bar \pi (t)+\hat \pi\,.
 \ee
The action of quadratic fluctuations for $\hat \pi$ around the scalar background  satisfies a linearised version
of symmetry \eqref{scalsym2}, i.e.
\be \label{scalsym3}
\delta \hat \pi\,=\,\bar \pi(t)\,\omega^a \partial_a \hat \pi\,.
\ee 
But such residual symmetry prevents the existence of a term quadratic in time derivatives in the quadratic action for $ \hat \pi$. Such term would violate the residual symmetry \eqref{scalsym3}, since the former is not invariant under the latter:
\bea
\delta\,\,\int d^4 x\, \left( \frac12 \dot{ \hat \pi}^2\right)&=&\int d^4 x\, \dot{ \hat \pi} \partial_t \delta { \hat \pi}
\nonumber \\
&=&\int d^4 x\, \dot{ \hat \pi}\left( c_0\, \omega^a \partial_a \hat \pi +  c_0 t\, \omega^a \partial_a \dot{\hat \pi}\right)
\nonumber \\
&=&c_0\, \int d^4 x\, \dot{ \hat \pi} \omega^a \partial_a \hat \pi
\,\neq\,0\,.
\eea
In the last step we integrated by parts, and removed a total derivative. This fact shows that, around flat space and a time-like scalar background, scalar excitations are non-dynamical since a symmetry prevents them to  acquire standard  kinetic terms.  We identified a decoupling limit where a new symmetry arises, and the action \eqref{actdec} propagates only four degrees of freedom, two tensors and two vectors.

\subsection{A symmetry outside the decoupling limit
}\label{sym-out}

It is natural to ask whether the results
of Section \ref{sym-dec} 
 remain valid also outside
a decoupling limit,
 in particular whether the theory described by Lagrangians \eqref{LagCur} propagates four or less degrees of freedom.
  The analysis in this case is made more subtle by the fact that the theory
 spontaneously breaks Lorentz symmetry, due to the presence of non-analytic functions 
 in the formulation of the action. A proper Hamiltonian analysis would require the  classification of  primary
 and secondary constraints, but this goes beyond the scope of this work. For studies of related problems in the context of scalar-tensor
 theories, see e.g. \cite{Chen:2012au,notensors}. 
  
  Nevertheless, again using symmetry arguments, 
we are able to provide indications 
 that  our systems propagate less than the expected  five {\it dofs}, 
 at least for 
some of the Lagrangians of eqs \eqref{LagCur}, and around certain backgrounds.
 We impose a discrete parity symmetry to the vector-tensor theories
we consider 
 \be
 A_\mu\,\to\,- A_\mu\,,
 \ee 
which singles out Lagrangians  $L_2$ and
 $L_4$ from the system of eqs \eqref{LagCur}:
 \bea \label{out2}
L_2&=& m^3\,\sqrt{-g}\,\sqrt{X}\,,
\\  \label{out4}
L_4&=& m\,\sqrt{-g}\,\left[
 \sqrt{X} R +  \frac{1}{2 \,\sqrt{X}}\left[(\nabla_\mu A^\mu)^2 - \nabla_\rho A_\sigma \nabla^\sigma A^\rho   \right]
\right]\,.
\eea
 These Lagrangians  are particularly important for the analysis of spherically symmetric configurations describing compact objects, as we shall discuss in the next section. 
 
 Lagrangian $L_{(2)}$ of eq \eqref{out2} and a  modification of $L_{(4)}$ given by

 \be
 \tilde{L}_{(4)}\,=\, m\,\sqrt{-g}\,\left[
 \sqrt{X} R +  \frac{1}{2 \,\sqrt{X}}\left[(\nabla_\mu A^\mu)^2 - \nabla_\rho A_\sigma \nabla^\sigma A^\rho-\frac14\,F_{\mu\nu}^2   \right]
\right]
 \label{out4as}
 \ee
are independently invariant under a  new gauge symmetry acting on the metric only,
\bea \label{newsez}
g_{\mu\nu}&\to&g_{\mu\nu}+ \partial_\mu \xi\,A_{\nu}+ A_{\mu}\,\partial_\nu \xi\,,
\eea
for arbitrary scalar function $\xi$. We checked this statement by brute force using the Mathematica package xAct \cite{xAct}.
  In Appendix  \ref{app-A}, we explain the heuristic method we used to deduce the existence of this symmetry: it arises
  from a certain limit of a disformal transformation acting on theories equipped by standard Abelian gauge invariance.

It would be interesting to derive in full generality the consequences of this new symmetry, in particular for what respect the total number
of degrees of freedom which it allows to propagate \footnote{In particular, it would be interesting  to understand whether this symmetry -- in the the scalar-tensor case where $A_\mu\,=\,\partial_\mu \phi$ -- provides a deeper reason for the fact 
that the scalar-tensor versions of 
 $L_{(2),\,(4)}$ do not propagate tensor modes, see \cite{notensors}.}. Here we focus our attention to the dynamics of fluctuations around flat space: we support the findings of Section \ref{sym-dec} but this time taking into full account the coupling between gravity and the vector longitudinal mode.
 
 We `switch off' the vector transverse modes, and only focus on the vector longitudinal component, which is a scalar whose dynamics
 we wish to study:
 \be
 A_\mu\,=\,\partial_\mu \pi\,.
 \ee
 In this case, Lagrangian $L_{(2)}$ reduces to   the cuscuton model, which has been shown not to propagate scalar degrees of freedom \cite{Afshordi:2007yx}: hence
 we do not consider it here any further.
The  
 two Lagrangians $L_{(4)}$ and $\tilde L_{(4)}$ of eqs \eqref{out4} and \eqref{out4as}  coincide in this scalar-only limit, and give the scalar Lagrangian
 \be
 L_{(4)}^{s}\,=\,m\,\sqrt{-g}\,
 \left[
 \sqrt{X_s} R +  \frac{1}{2 \,\sqrt{X_s}}\left[(\nabla_\mu \partial^\mu\,\pi)^2 - \nabla_\rho \partial_\sigma \pi\, \nabla^\sigma \partial^\rho \pi  
  \right]
  \right], 
 \ee
 which describes interactions of the vector longitudinal mode with gravity and with itself. This scalar Lagrangian is invariant under 
 a scalar-tensor limit of symmetry \eqref{newsez}:
 \bea \label{newsez2}
 g_{\mu\nu}&\to&g_{\mu\nu}+ \partial_\mu \xi\,\partial_{\nu} \pi+ \partial_{\mu} \pi\,\partial_\nu \xi\,,
 \eea
 for arbitrary  $\xi$.
 
  We now add an Einstein-Hilbert term to the system and analyse the action
 \be \label{actS1a}
S\,=\,\int d^4 x \sqrt{-g} \left[ 
M_{Pl}^2 R
+ \lambda_4 \,L_{(4)}^{s}
 \right]\,,
\ee   
describing  gravity coupled with the vector longitudinal mode. We show that symmetries available -- standard diffeomorphisms, as well as the new symmetry 
 \eqref{newsez2} 
-- prevent the propagation of the scalar excitation around flat space. 
Notice that the analysis is made more subtle by the fact that, while $L_{(4)}^{s}$ respects the new symmetry
 \eqref{newsez2}, the Einstein-Hilbert term is in general {\it not} invariant under this symmetry (although it might
 be in specific cases). 

Flat space $g_{\mu\nu}\,=\,\eta_{\mu\nu}$ is a background solution for the equations of motion associated with the  action \eqref{actS1a}, for any constant time-like vector background
\be \label{scalbt}
\bar \pi\, \equiv\,\bar c_\mu x^\mu\,.
\ee
  We study the dynamics of fluctuations around this background.
  Perturbations are denoted as
\bea
g_{\mu\nu}&=&\eta_{\mu \nu}+ \hat h_{\mu\nu}\,,
\\
\pi&=&\bar \pi+ \hat \pi\,,
\eea
with $ \hat h_{\mu\nu}$ metric fluctuations, and  $\hat \pi$ 
the scalar fluctuations.

Local infinitesimal diffeomorphisms
 act on metric and scalar  fluctuations as
\bea \label{dift1}
\hat h_{\mu\nu}&\to& \hat h_{\mu\nu}+\partial_\mu\,\alpha_\nu+\partial_\nu\,\alpha_\mu\,,
\\
\label{dift2}
\hat \pi&\to&\hat \pi+ \bar c_\rho  \alpha^\rho\,,
\eea
for an arbitrary  function $\alpha^\mu$ of small size, with $\bar c_\rho$ the vector controlling the 
 scalar background \eqref{scalbt}. A non-vanishing background for $\bar \pi$ spontaneously breaks
diffeomorphisms (introducing the last contribution to eq \eqref{dift2}). 
 Symmetry
 \eqref{newsez2}, distinctive of  the  $L_{(4)}^s$ contribution to action  \eqref{actS1a},  gets also spontaneously broken by the scalar  $vev$. It acts on the metric fluctuations as
  \bea \label{difdis1}
\hat h_{\mu\nu}&\to& \hat h_{\mu\nu}+ \bar c_\mu \,\partial_\nu \xi+ \bar c_\nu \,\partial_\mu \xi
 \eea 
 for an arbitrary quantity $\xi$ of small size. Notice that  symmetry \eqref{difdis1}  is equivalent  
 to  a linearised diffeomorphism acting on the metric only, hence in this case the Einstein-Hilbert term is invariant under this transformation
 as well.

To analyse the dynamics of scalar fluctuations, 
we  select a gauge and choose a convenient profile for the diffeomorphism parameter  $\alpha^\mu$:
 \bea
 \alpha^\mu&=&-\left(
\frac{\bar c^\mu}{\bar c^\rho \bar c_\rho}\right)\,\hat \pi\,.
 \eea
Such  unitary gauge  moves the   fluctuation $\hat \pi$ from the scalar to the metric fluctuations, that gets 
a contribution depending on derivatives of  $\hat \pi$
  (see eq \eqref{dift1}):
  \bea \label{dift3}
\hat h_{\mu\nu}-
\frac{\bar c_\mu \,\partial_\nu\hat \pi}{\bar c^\rho \bar c_\rho}
-
\frac{\bar c_\nu \,\partial_\mu\hat \pi}{\bar c^\rho \bar c_\rho}\,.
\eea
In principle,  $\hat \pi$ might acquire dynamics when  expanding  $L_{(4)}^s$
  quadratically in fluctuations.
   However, we can still exploit  the symmetry \eqref{difdis1}, to show that 
    $\hat \pi$  is a non-dynamical gauge mode also for  $L_{(4)}^s$.
   Making the
choice 
 \bea
\xi^\mu
&=&\frac{\partial^\mu \hat \pi}{\bar c^\rho \bar c_\rho}
\eea
in eq \eqref{difdis1}
we can remove the $\pi$ contribution from \eqref{dift3}. 
 Hence, around flat space, the scalar  mode does not propagate.

  \smallskip

It would be interesting to further exploit consequences of symmetry \eqref{newsez} to study in full detail  the dynamics of fluctuations
also around more general background configurations, and including transverse vector modes. As the brief analysis
above shows, the number of propagating degrees of freedom should be controlled on the symmetries available
around a given configuration.
We leave this issue for the future. 
  In the next Section we are going to examine    the physics of spherically 
 symmetric and slowly rotating  solutions to the equations of motion associated with  specific vector-tensor
  actions.
 
\section{Spherically symmetric  and slowly rotating solutions}\label{sec-solg}

In the previous Section, we discussed 
  candidates of vector-tensor theories
  that can propagate less than the expected
five degrees of freedom, even in absence of an Abelian gauge symmetry. Vector-tensor theories
of gravity  that do not propagate the vector longitudinal mode
can be interesting for phenomenology, since they  automatically avoid long range fifth forces associated with light scalar
fields. 
   In this Section, we discuss
    ramifications of these theories for configurations that can describe compact objects.
 We   
         focus on a  subset of these actions which obey a further parity symmetry $A_\mu\to-A_\mu$: the Lagrangians  $L_{(2),\,(4)}$ given by
 eqs \eqref{out2}, \eqref{out4},
  \bea \label{out2a}
L_{(2)}&=& m^3\,\sqrt{-g}\,\sqrt{X}\,,
\\  \label{out4a}
L_{(4)}&=& m\,\sqrt{-g}\,\left[
 \sqrt{X} R +  \frac{1}{2 \,\sqrt{X}}\left[(\nabla_\mu A^\mu)^2 - \nabla_\rho A_\sigma \nabla^\sigma A^\rho   \right]
\right]\,.
\eea
  We   examine
      spherically symmetric and slowly rotating 
    configurations associated with such Lagrangians.

    In Section \ref{Sec-stealth} we consider an action which includes the Einstein-Hilbert contribution, plus Lagrangians $L_{(2)}$ and  $L_{(4)}$  of eqs \eqref{out2a}, \eqref{out4a}.
    We prove a  theorem on the existence of   stealth configurations  in the 
     presence of 
      arbitrary additional matter besides the vector-tensor Lagrangians we consider. The resulting regular
      solutions 
      have the same geometry as in Einstein gravity coupled with matter, and 
       a non-trivial vector profile       characterized by independent vector  charges. 
       We also consider physical consequences
       of  our findings in terms of the degrees of freedom the system can propagate, making further connections with the results
       of Section \ref{sym-out}. These results can be important to describe configurations describing black holes or neutron stars in
        modified theories of gravity including more realistic astrophysical  matter forming or surrounding the object. 
       
       In Section \ref{Sec-max} we add to the action 
         the standard Maxwell kinetic term for the vector fields in the action, and we show that 
       spherically   symmetric configurations correspond to a Reissner-Nordstr\"om configuration, plus small subleading corrections at large radial distances. We discuss how the geometry depends on features  of Lagrangians
       $L_{(2)}$ and  $L_{(4)}$.

\subsection{Stealth 
  spherically symmetric and slowly rotating configurations } \label{Sec-stealth}

Consider the  action
\be \label{actma1}
S\,=\,\int d^4 x\,\sqrt{-g}\,\left[ M_{Pl}^2\,R+\lambda_2 L_{(2)}+\lambda_4 L_{(4)}+{\cal L}_{matter}\right]\,,
\ee
with $L_{(2), (4)}$ the vector-tensor Lagrangian densitites of eqs \eqref{out2a}, \eqref{out4a}. 
${\cal L}_{matter}$ describes an {\it arbitrary} matter
Lagrangian, with the only condition that matter does not directly couple with  the vector field $A_\mu$
 (but only indirectly through gravity). We prove that the system admits   spherically symmetric solutions  -- as well as solutions in slow rotation -- 
which coincide  exactly with the ones of Einstein gravity coupled with ${\cal L}_{matter}$ 
 (i.e. with the solutions obtained without the  $L_{(2), (4)}$ contribution to the previous action). 
  These are examples of stealth spherically symmetric
configurations  
  with 
 a non-trivial profile for the vector field. 

\subsubsection*{The general proof}


The metric Ansatz we adopt for a general spherically symmetric configuration is 
\be \label{metans1}
d s^2\,=\,-F(r) \,d t^2+2 \,D(r)\,d t \,d r\,+H(r)\,d r^2+r^2\,d \Omega^2\,,
\ee
We choose
 an Ansatz for the gauge field profile preserving the spherical symmetry, and with only the time-component turned on 
 \be\label{ansvec1}
A_\mu\,=\,\left( A_0(r),\,0\,,0\,,0\right)\,.
\ee

When considering spherically symmetric configurations and the metric  Ansatz \eqref{metans1},
 the radial profile of the metric component $D(r)$ {\it or} of $H(r)$   can be
 chosen arbitrarily by means of a coordinate redefinition. Then the remaining metric components are uniquely determined by the
equations of motion. 
 Indeed, 
a shift of time coordinate
\be \label{tish1}
d t\,\to\,d t+\frac{G(r)}{F(r)}\,d r\,,
\ee
for some arbitrary function $G(r)$ leads to the following metric and vector field profiles:
\be \label{metans2}
d \tilde{s}^2\,=\,-F(r) \,d t^2+2\left( D(r)-G(r) \right)\,d t dr +\left( H(r) +\frac{G^2(r)}{F(r)} \right)\,d r^2+r^2\,d \Omega^2\,,
\ee
 \be \label{ansvec2}
\tilde A_\mu\,=\,\left( A_0(r),\,\frac{G(r)\, A_0(r)}{F(r)}\, ,0\,,0\right)\,.
\ee
Usually  such coordinate freedom is used  
  to remove the
   off-diagonal component,  making
   the specific  choice  
%
$G=D$ in eq \eqref{metans2}.  On the other hand, in what comes next we need  to have gauge freedom to choose a profile for $H(r)$, hence
we maintain  the general Ansatz \eqref{metans1} including the off-diagonal component.

\smallskip

With the metric and field Ansatz \eqref{metans1}, \eqref{ansvec1}, we proceed to study the equations of motion. 
There are four equations to satisfy,  associated with the four quantities $A_0(r)$, $F(r)$, $D(r)$, $H(r)$.

Plugging the Ansatz
in eq \eqref{actma1}, we find that $A_0(r)$ appears only linearly in the action, and multiplies a combination depending only on $H(r)$. 
This property is distinctive of our choice for $G_{2,\,4}$ in eqs \eqref{g2h} and \eqref{g4h}.
The action
results  
\bea \label{actA01}
S&=&4 \pi\,m\,\int d r \,
A_0(r)\,H^{-3/2}(r)
\,\left[
-\sqrt{2} \lambda_4\,H(r)+\left(m^2\, r^2 \lambda_2+\sqrt{2} \lambda_4\right) H^2(r)+\sqrt{2}\,\lambda_4 \,r\,H'(r)
\right]
\nonumber
\\
&&+ {\text{  \,\,parts that do not depend on $A_0$,}}
\eea
with the prime denoting a derivative along the radial coordinate. 
Hence
 the vector
equation of motion for the component $A_0(r)$ is special, since it provides a constraint condition on the geometry:
\be
0\,=\,-\sqrt{2} \lambda_4\,H+\left(m^2\, r^2 \lambda_2+\sqrt{2} \lambda_4\right) H^2+\sqrt{2}\,\lambda_4 \,r\,H'\,,
\ee
with solution
\be \label{proH1}
H(r)\,=\,\left( 1-\frac{2 \mu}{M_{Pl}^2\,r}+\frac{m^2\,\lambda_2\,r^2}{6 \,\lambda_4}\right)^{-1}\,,
\ee
where $\mu$ is an arbitrary integration constant with dimension of a mass.  Recall that, as discussed around \eqref{metans2},
  the    metric function $H(r)$ is a `pure gauge' and its profile  can be modified 
by changing coordinates. But for the moment we work with the solution \eqref{proH1}
for the vector field equation, and proceed to characterise the remaining unknown quantities, determined by the equations of motion associated with the metric components. 

The equations of motion for the metric components
$F(r)$, $D(r)$ are  insensitive to the vector component $A_0(r)$ (since the dependence of the action on $A_0(r)$ is limited
to the first line of eq \eqref{actA01}), hence they have the very same solutions as in Einstein gravity coupled with matter Lagrangian
${\cal L}_{matter}$ (in other words, they do not realise the presence of $L_{(2), (4)}$ in eq \eqref{actma1}).

Finally, the equation of motion relative to $H(r)$ 
depends explicitly on the vector profile,
and
 can be used for determining
  $A_0 (r)$.
  The vector field profile 
 is non-trivial, and is controlled by the geometry (without affecting it).
 In general, we expect that the vector profile depends on two independent integration constants (besides the ones that control the geometry): one is the integration
 constant $\mu$ entering in the solution of the eom for $A_0(r)$
  in eq \eqref{proH1}. The other an integration constant associated with the eom for $H(r)$ (we will expand on  this statement in
 a specific example next).
  After
   having determined the solution,
  if one wishes, one can make a coordinate transformation as in eq \eqref{metans2},  setting the metric
in a diagonal form choosing $G(r)=D(r)$,  and turning on a radial component for the vector profile.  From the
arguments above, the geometry is the one of Einstein gravity coupled with ${\cal L}_{matter}$.

 \bigskip
 
 The very same general proof we have developed can be applied to describe configurations in slow-rotation, described by  the line element \cite{Carroll:2004st}
 
 \be \label{metans2}
d s^2\,=\,-F(r) \,d t^2+2 \,D(r)\,d t \,d r\,+H(r)\,d r^2+r^2\,d \Omega^2+2\,a\, r^2\,\sin^2{\theta}\,W(r)\,d\phi\, dt\,,
\ee
and
 \be\label{ansvec2}
A_\mu\,=\,\left( A_0(r),\,0\,,0\,,0\right)\,,
\ee
 with $a$ the rotation parameter.
 The corresponding system  of equations can be studied in an expansion at first order in $a$, leading to the very same conclusions we derived above.
When ${\cal L}_{m}=0$,  the function $W(r)$ acquires a profile $W\sim 1/r^3$, the same as in the slow rotation limit of the Kerr configuration in Einstein GR.  In presence of matter, the eom for $W(r)$ can be sourced by it. 
 
 \smallskip
 
 After  this general proof, we now discuss the case with cosmological constant
  as an  example to concretely explain  the structure of the
   solution,
      and to show the existence of two integration constants associated with the vector profile. 
 
\subsubsection*{A concrete example: space-time in the presence of  a cosmological constant}

It is well known (see e.g. the papers \cite{Rinaldi:2012vy}, and \cite{Volkov:2016ehx} for a general review) that adding a cosmological constant to generic
  Horndeski scalar-tensor  systems lead to spherically symmetric black hole solutions which are quite different from
  their counterparts in Einstein gravity i.e. (A)dS-Schwarzschild black holes. On the other hand, for the specific choice of vector-tensor action \eqref{actma1}
  with ${\cal L}_{matter}\,=\,-6 \,\Lambda$,   our arguments suggest that there is a solution to the field equations whose geometry
   coincides with the one of General Relativity. 
  
  We can prove this fact directly, proceeding
  as discussed above. 
  Adopting a spherically symmetric metric and vector field Ansatz as in eqs  \eqref{metans1},
   \eqref{ansvec1} 
   with an off-diagonal component for the metric, 
   it is straightforward to find the explicit, unique solution:
\bea
H(r)&=&\left( 1-\frac{2 \mu}{M_{Pl}^2\,r}+\frac{m^2\,\lambda_2\,r^2}{6 \,\lambda_4}\right)^{-1}\,,
\\
F(r)&=&1-\frac{2 M}{M_{Pl}^2\,r}-\frac{\Lambda\,r^2}{M_{Pl}^2}\,,
\\
D^2(r)&=&\frac{6\,\lambda_4\,\Lambda\,r^3+\lambda_2\,m^2\,M_{Pl}^2 \,r^3-12\,\lambda_4\,\left(\mu-M \right) }{ \lambda_2\,m^2\,M_{Pl}^2  r^3-12
\lambda_4  \mu +6 \lambda_4 r}\,,\\
A_0(r)&=& {Q}\,\left(
\lambda_4-\frac{2\,\lambda_4\,\mu}{M_{Pl}^2\,r}
+\frac{\lambda_2\,m^2\,r^2}{6}
\right)^{1/2}\,.
\eea
This configuration depends on {\it three} integration constants: the parameter $\mu$ (with dimension of mass), which enters in the profile of $H(r)$, and is associated with the eom for $A_0(r)$;
the  mass parameter $M$,  which enters in the profile of $F(r)$, and comes from solving the equations of motion for $D(r)$, $F(r)$; and finally, the parameter $Q$
in the profile for $A_0(r)$ (with dimension of mass) associated with the eom for $H(r)$. 

By doing a coordinate redefinition as explained around eq \eqref{metans2}, the metric can be put in a  diagonal form, which makes it  more recognizable:
\be
d s^2\,=\,-\left(1-\frac{2 M}{r}-\frac{\Lambda\,r^2}{M_{Pl}^2}\right)\,d t^2+\frac{d r^2}{\left(1-\frac{2 M}{r}-\frac{\Lambda\,r^2}{M_{Pl}^2}\right)}
+r^2\,d \Omega^2 \,.
\ee
As expected, this is  a stealth (A)dS Schwarzschild black hole configuration, which depends on the integration constant $M$, the mass of the 
black hole. On the other hand, in this frame,
 the vector profile is non-trivial, and has two components switchted on: 
\bea
A_\mu&=& \left( A_0(r),\,\Pi(r),\,0,\,0\right)\,,
\\
A_0(r)&=& {Q}\left(
\lambda_4-\frac{2\,\lambda_4\,\mu}{M_{Pl}^2\,r}
+\frac{\lambda_2\,m^2\,r^2}{6}
\right)^{1/2}\,,
\\
\Pi^2(r)&=&\frac{m\,Q^2}{M_{Pl}^2\,r} \frac{\left[
\lambda_2\,
m^2\,M_{Pl}^2 r^3
+6\,\lambda_4\,\left(
2 M+r^3\,\Lambda-2 \mu
\right)
 \right]^{1/2}}{\left(1-\frac{2 M}{M_{Pl}^2\,r}+\frac{r^2\,\Lambda}{M_{Pl}^2} \right)}\,.
\eea
The vector field profile  contains  two integration constants: $Q$, which can be interpreted as the standard vector `electric' charge associated with the vector
component $A_0$; and $\mu$, a new parameter that contributes to characterize the vector solution. 

\subsubsection*{Physical consequences}

These results
 ensure that there exist spherically symmetric and slowly rotating solutions 
of action \eqref{actma1} which coincide with the ones of Einstein gravity, coupled with matter Lagrangian ${\cal L}_{matter}$. At the same time, the vector field acquires a non-trivial profile, which depends on the geometry and extends
 to asymptotic infinity and is characterized by specific, {independent} integration constants. These findings are useful for embedding these vector-tensor theories of gravity in more
 realistic settings, including astrophysical matter forming or surrounding compact, gravitationally bound objects.
  
  It is tempting to relate the findings of this  Section to the results of Sections \ref{sym-dec} and  \ref{sym-out}, which indicate that the theories
  under consideration propagate less than the expected degrees of freedom: possibly,  spherically symmetric configurations coincide with the ones of Einstein gravity, because matter in this case does not excite additional vectorial degrees of freedom, which can backreact on the geometry modifying it. It would be interesting to further pursue these arguments to understand how the conditions of finding stealth configurations in modified gravity theories are associated with the number of propagating modes. 
  
  It is also interesting to ask how to probe the non-trivial, asymptotic vector profile associated with the  geometry, and the  vector `hairs' associated with the specific vector integration constants. A possibility is to couple the vector to probe matter fields, although then backreaction effects of  the vector to the geometry should be taken into account. Alternatively, we can give kinetic terms to the vector in the form of a standard Maxwell action: we discuss this case in what follows.

\subsection{Including Maxwell action: small deviations from Reissner-Nordstr\"om }\label{Sec-max}

We now investigate how the results of the previous Section are modified when including vector kinetic terms to the 
system. 
We consider the action
\be
S\,=\,\int d^4 x\,\sqrt{-g}\,\left[ M_{Pl}^2\,R-\frac14\,F_{\mu\nu} F^{\mu\nu}+\lambda_2 L_{(2)}+\lambda_4 L_{(4)}\right],
\ee
with the choices \eqref{out2a} and \eqref{out4a} for  $ L_{(2), \,(4)}$, and without loss of generality we set $m=M_{Pl}$.
We  find that spherically symmetric solutions, in absence of additional matter, resemble Reissner-Nordstr\"om (RN) configurations {in certain regimes}, with small deviations induced by the vector-tensor Lagrangians $ L_{(2), \,(4)}$.

The Ansatz for the metric and the gauge field we start with is the same as in \eqref{metans1}.
When the vector kinetic term is absent, $\lambda_2$ drives the vector field 
profile to diverge asymptotically, although the metric remains asymptotically flat. 
When the vector kinetic term is included, $\lambda_2$ spoils the asymptotic flatness 
of the metric. Therefore,  we start with the simpler case $\lambda_2=0$.
The metric components $F$, $H$ and $D$ can be written in terms of $A_0$ and one integration constant that we call $M$. $A_0(r)$ satisfies the equation
\begin{equation}
\frac{2 \sqrt{2} \lambda _4 \left(A_0 \left(r A_0''+2 A_0'\right)+r \left(A_0'\right){}^2\right)}{ \sqrt{A_0 \left(2 r A_0'+A_0\right)^3}}-  \left(r A_0''+2 A_0'\right)=0\,. \label{eq:A0}
\end{equation}
At large distances, this equation admits a parametric solution in terms of an inverse radius  expansion
\begin{equation}
A_0 = P + \frac{Q}{r} + \sum_{i=2} \frac{a_i}{r^i}\,,
\end{equation} 
where the coefficients $a_i$ are determined by $P$, $Q$ and the parameters of the model.  $M$ does
not appear in these coefficients. For example, the first few values of $a_i$ are
\begin{align}
a_2 &= \frac{Q^2 \lambda _4 \left(\sqrt{2} P+4 \lambda _4\right)}{P^3 -8 P \lambda _4^2}\,,\\
a_3 & = \frac{Q^3 \lambda _4 \left(\sqrt{2} P^3 +20 P^2  \lambda _4+56 \sqrt{2} P  \lambda _4^2+96 \lambda _4^3\right)}{3 \left(P^3-8 P \lambda _4^2\right){}^2}\,, \\
a_4 & = \frac{Q^4 \lambda _4 \left(3 \sqrt{2} P^5+88 P^4 \lambda _4+504 \sqrt{2} P^3  \lambda _4^2+2784 P^2 \lambda _4^3+3712 \sqrt{2} P \lambda _4^4+3840 \lambda _4^5\right)}{12 \left(P^3 -8 P \lambda _4^2\right){}^3}\,.
\end{align}
After diagonalising the metric as explained around eq \eqref{metans2}, we find
\begin{align}
f & = h = 1-\frac{2 M}{r}+\frac{Q^2 }{4 r^2}+\frac{Q^3 \lambda _4 \left(\sqrt{2} P+4 \lambda _4\right)}{2 r^3 \left(P^3 -8 P \lambda _4^2\right)} 
+ \dots ,\\
\Pi & = \sqrt{\frac{2P (M P+Q)}{r} } + \frac{8 P (2 M P+Q)^2 -P^3 Q^2-256 M (M P+Q) \lambda _4^2+8 Q^2 \lambda _4 \left(2 \sqrt{2}+P \lambda _4\right)}{8 \sqrt{2( M+ Q/P)} r^{3/2}\left(P^2-8 \lambda _4^2\right)} + \dots
\end{align}
This shows that, at large $r$, the solutions correspond to small deviations from a RN black hole, with a non-trivial profile for the longitudinal mode
of the vector field. 

To investigate what happens for small $r$ we resort to a numerical analysis. We solve eq. \eqref{eq:A0} by imposing initial conditions
for $A_0$ and $A_0'$ at a large radius $r_i$ and evolve inwards. We verify that $r_i$ is large enough for the solutions to respect
the asymptotic behaviour found analytically, so that we can safely read from the solutions the black hole mass $c^2 M/G$, and the vector charges $P$ and $Q$. We determine the radius of the black hole horizon $r_h$ by identifying the point where $g^{rr}=0$. 
\begin{figure}
\includegraphics[width=0.48\textwidth]{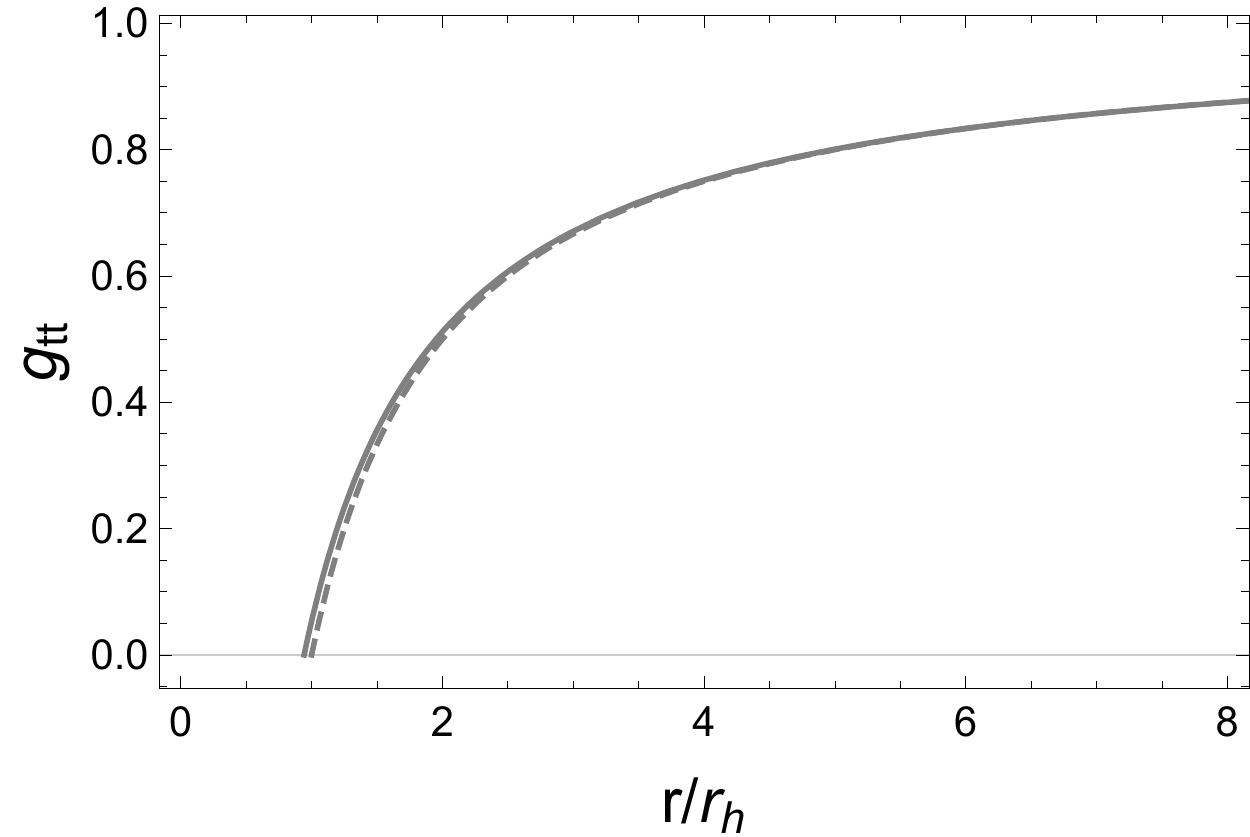} \ \includegraphics[width=0.48\textwidth]{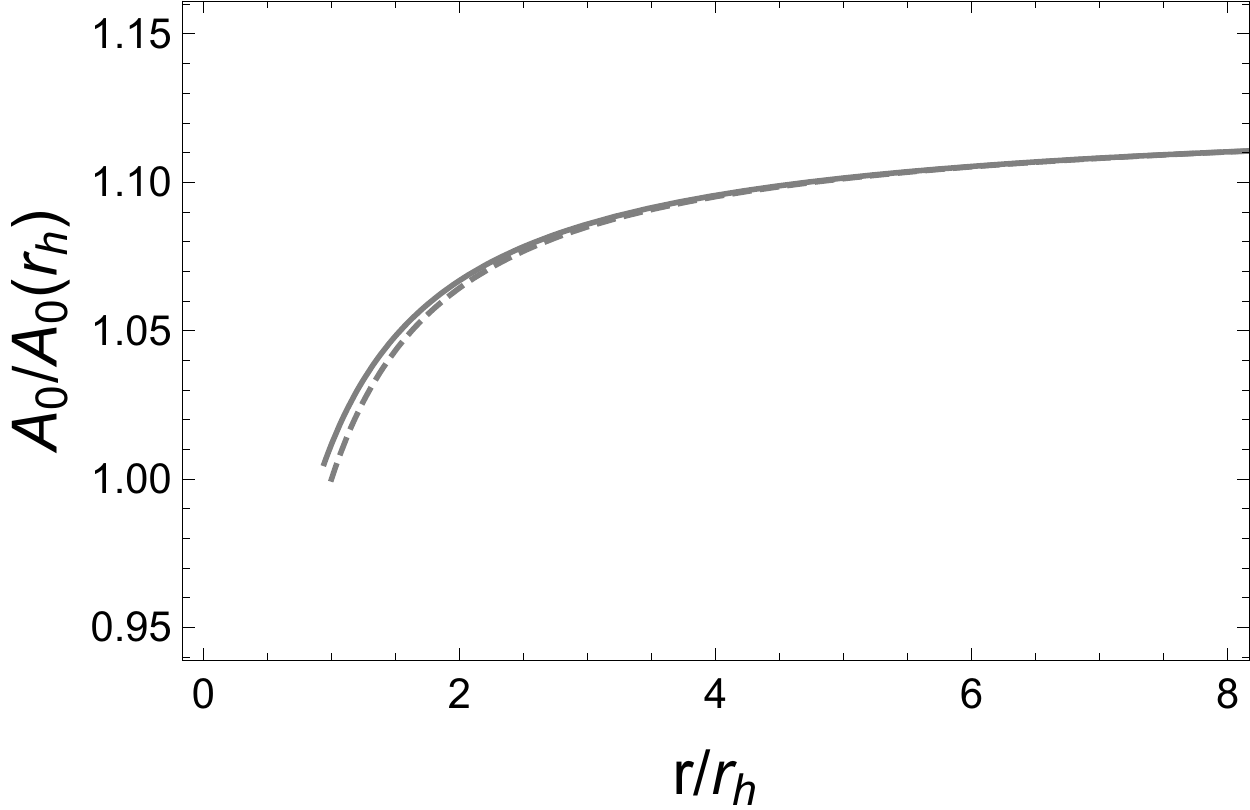} 
\caption{$g^{tt}$ and $A_0$ for $\lambda_2 = 0$. The dashed line shows the case without vector kinetic term, where  Schwarzschild BH's  of radius $r_h$ are formed; for both solutions -- with and without vector kinetic term -- the radial coordinate is normalized over this $r_h$. When the Maxwell term is on, the horizon position is shifted to  smaller values. In both cases, the metric is asymptotically flat and $A_0$ asymptotes to a constant value. For these plots,
we use $\lambda_4 = 1$, and the asymptotic values $P \approx 5 M_p $ and $Q\approx -1600$ and mass $c^2M/G= 1500 M_p$ (in units where $c = G = M_p = 1$. Restoring the SI values of these constant, this is equivalent to 1 solar mass).  } 
\label{fig:l4}
\end{figure}
The results are shown in Fig.~\ref{fig:l4}. The left panel shows the profiles for $g_{tt}=g^{rr}$, and the right panel shows the profiles for $A_0$.
For comparison, the dashed lines show the case without 
 vector kinetic term, where the metric is exactly Schwarzschild. Only near the horizon the solution deviates from Schwarzschild. Interestingly, the resulting black hole horizon area is smaller than the one of a Schwarzschild 
or RN black hole with the same
mass. This implies that these black holes
are more compact than their GR or Einstein-Maxwell counterparts.  

\begin{figure}
\includegraphics[width=0.48\textwidth]{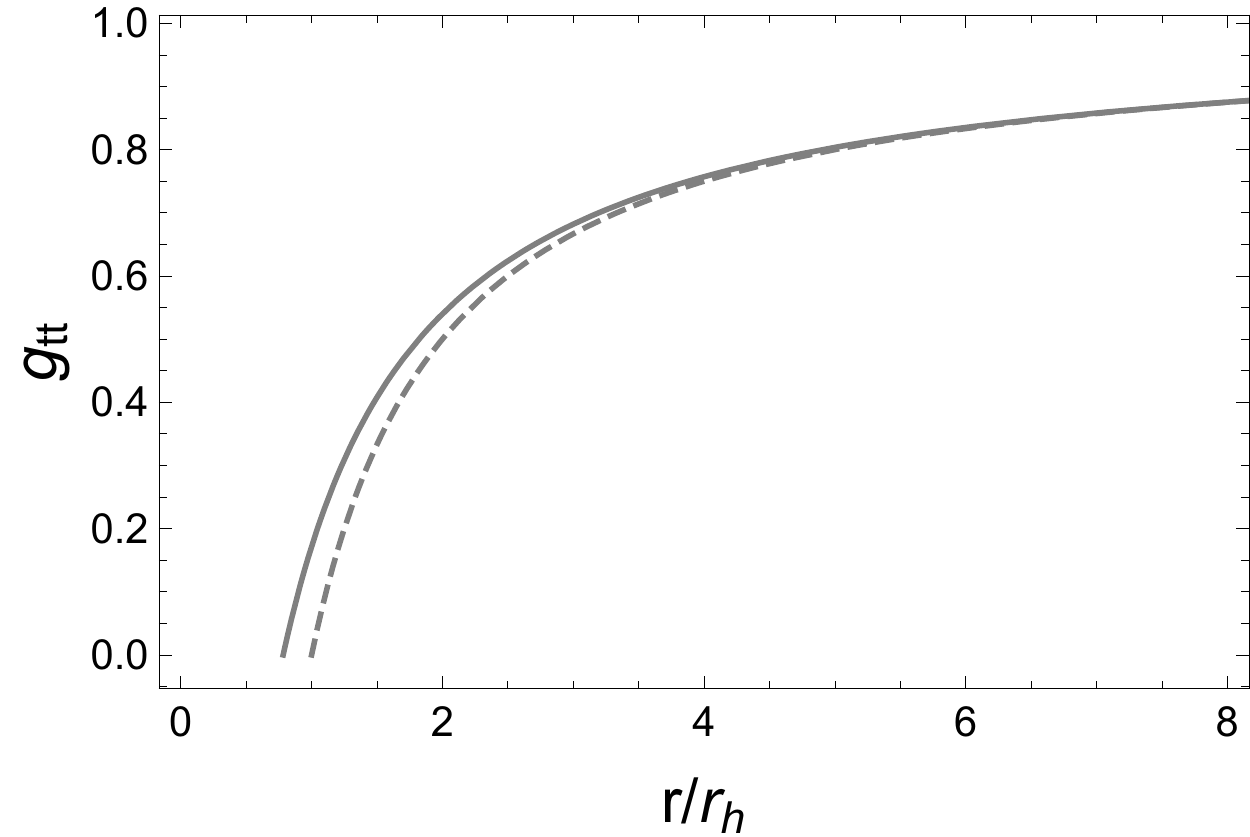} \ \includegraphics[width=0.48\textwidth]{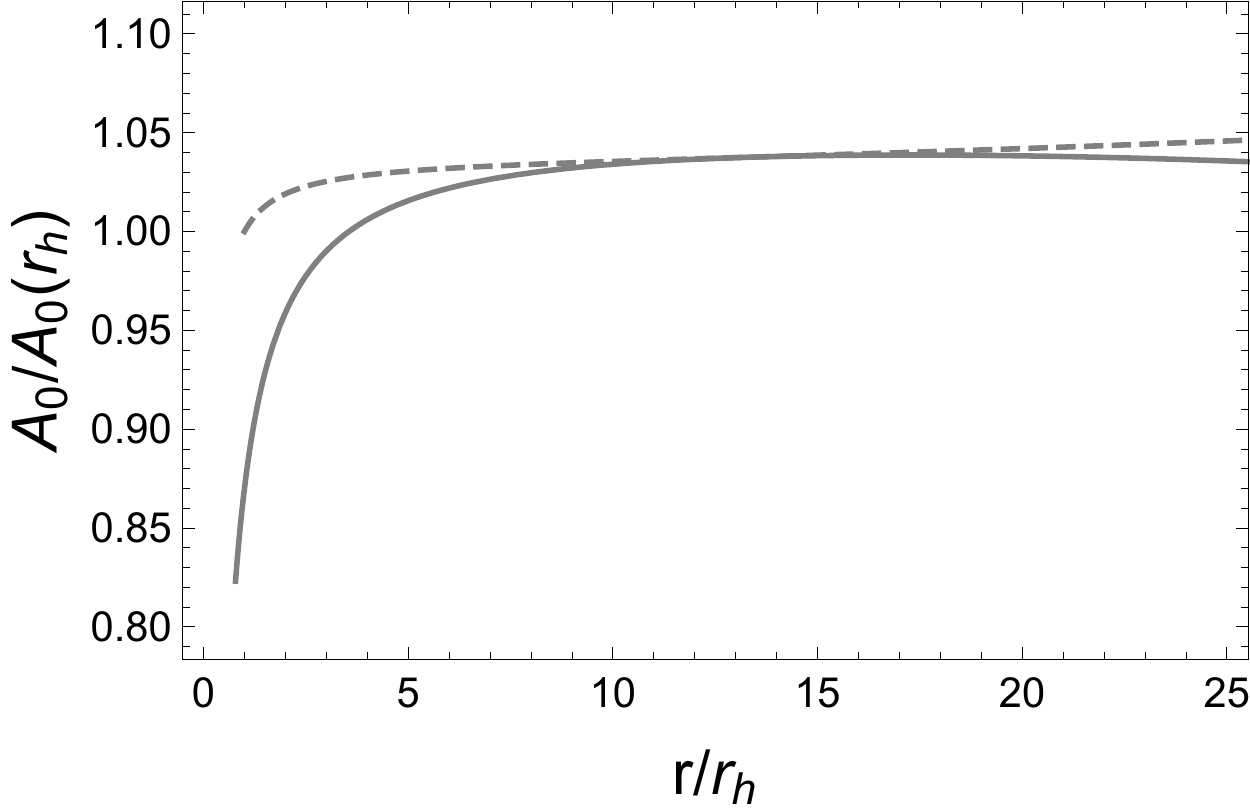}
\caption{Same as the previous figure but with both $\lambda_2$ and $\lambda_4$ different from zero. Only when the vector kinetic term is present
the asymptotic behaviour of the metric is sensitive to the asymptotic divergence of $A_0$. Since we do not have an analytic expression 
for the asymptotic solution we do not have control over the charges of this solution, therefore we set the same initial conditions as for the
case $\lambda_2 = 0$, and choose $\lambda_4 = 1$ and a small value $\lambda_2=2\times 10^{-11}$, such that these initial conditions are approximately valid.}
\label{fig:l4l2} 
\end{figure}

Now we turn our attention to the case where both $\lambda_2$ and $\lambda_4$ are different from zero\footnote{The case  $\lambda_4=0$, $\lambda_2\neq 0$ does not admit spherically symmetric solutions with a diagonal metric that satisfies $g_{tt}=g^{rr}$.}. Once again,
the metric components can be expressed in terms of $A_0$, which is determined from the equation
\begin{align}
\lambda _4 \left(2 \lambda _4+\lambda _2 r^2\right) \left(2 r A_0'+A_0\right){}^2-A_0 \left(2 \lambda _4+\lambda _2 r^2\right) \left(2 r A_0'+A_0\right) \left[r \sqrt{\frac{\lambda _4 \left(2 r A_0'+A_0\right)}{A_0 \left(2 \lambda _4+\lambda _2 r^2\right)}} \left(r A_0''+2 A_0'\right)\right.&\left.+\lambda _4\phantom{\frac12}\!\!\!\right]  \nonumber \\
- r \lambda _4 \left[\left(2 r A_0'+A_0\right) \left[A_0' \left(2 \lambda _4+\lambda _2 r^2\right)+2 A_0 \lambda _2 r\right]-A_0 \left(2 \lambda _4+\lambda _2 r^2\right) \left(2 r A_0''+3 A_0'\right)\right] &= 0 \nonumber
\end{align}
It is straightforward to verify that in the limit $\lambda_2 \to 0$ this reduces to eq. \eqref{eq:A0}. Since the solutions 
are not asymptotically flat, analytic expansions are difficult to implement, therefore we show only numerical results. To set 
initial conditions we assume a small value of $\lambda_2$, so that the same initial conditions used for the asymptotically flat solutions
with $\lambda_2=0$ can be used as an approximation.
 Fig.~\ref{fig:l4l2}
shows the metric component $g_{tt}$ and the vector component $A_0$. Both the metric and vector field components diverge for large $r$. 

\begin{figure}
\includegraphics[width=0.48\textwidth]{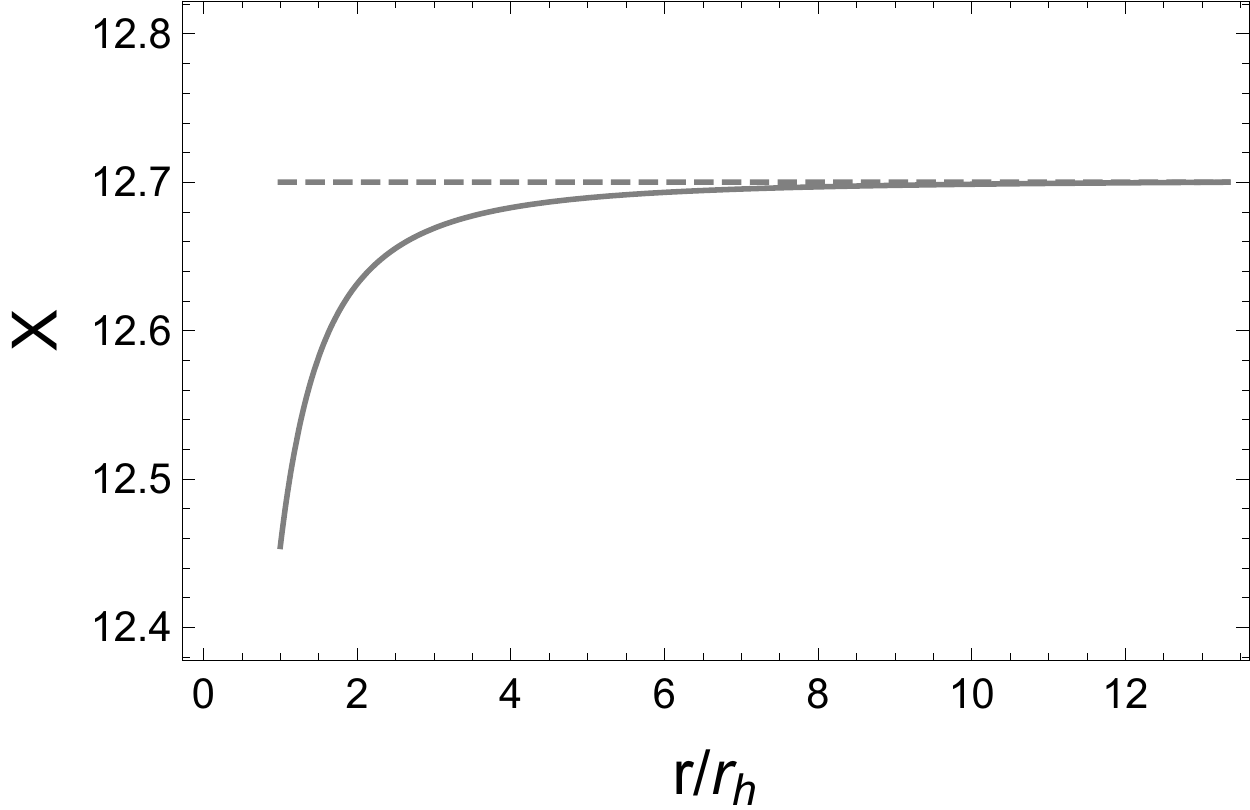} \ \includegraphics[width=0.48\textwidth]{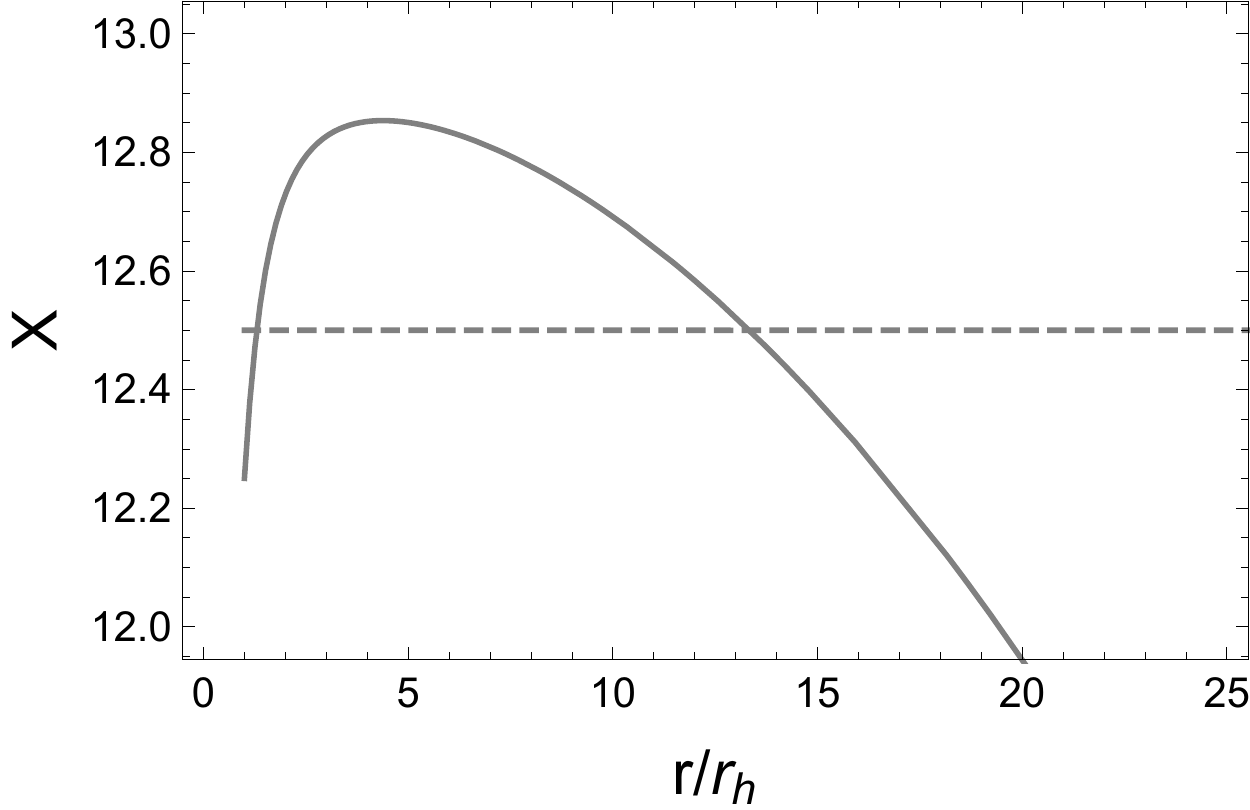}
\caption{ The left panel shows $X$ for $\lambda_2=0$ and $\lambda_4=1$. The right panel shows $X$ for $\lambda_2=2\times 10^{-11}$ and $\lambda_4=1$. In both panels, the dashed line shows the constant value
of $X$ obtained when the vector kinetic term is absent. }
\label{fig:x} 
\end{figure}

Finally, notice that the quantity $X= -\frac{1}{2} A_\mu A^\mu$ has a non-trivial profile, as shown in Fig.~\ref{fig:x}. This is a difference with respect to configurations with no vector kinetic terms.
For $\lambda_2=0$ we can  write down an asymptotic analytic expression for $X$. Up to second order in $1/r$, it has the form
\begin{equation}
X  = \frac{1}{2} A_0 (A_0 + 2 r A_0') = \frac{P^2}{2}+\frac{P Q^2 \left(2 \sqrt{2} \lambda _4+P\right)}{2 r^2 \left(8 \lambda_4^2-P^2\right)} + \dots 
\end{equation}

For $\lambda_2 \neq 0$, we can write an expression for $X$ in terms only of $A_0$,
\begin{equation}
X = \frac{A_0 \lambda _4 \left(2 r A_0'+A_0\right)}{2 \lambda _4+\lambda _2 r^2},
\end{equation} 
but we cannot approximate its behaviour analytically since we do not know the asymptotic profile of $A_0$.
\smallskip

To conclude, in this Section we shown that when including vector kinetic terms, the geometry depends on the details of the vector profile: when $\lambda_2=0$, the geometrical solution corresponds to a small deformation from RN black hole. When $\lambda_2\neq0$, the vector backreaction is more important, and the geometry is no more asymptotically flat. 

\section{Discussion}\label{sec-concl}
 In this work, we have shown that specific examples of 
  vector-tensor  theories of gravity have novel gauge symmetries, which can prevent the propagation  
  of  the vector longitudinal degree of  freedom, at least around certain backgrounds. Vector-tensor theories
of gravity  that do not propagate the vector longitudinal mode
can be interesting for phenomenology, since they  automatically avoid long range fifth forces associated with light scalar
fields. We provided an heuristic understanding of these symmetries, as originating from a vector disformal transformation
of an Abelian symmetric set-up. 
  We then studied properties of spherically symmetric and slowly rotating configurations in this system. We
 have shown that a system containing Einstein gravity, vector-tensor Galileons, and an  arbitrary matter Lagrangian (with no direct couplings to vectors) admits stealth solutions 
  whose geometry coincides with solutions of Einstein gravity coupled with the  matter Lagrangian. Nevertheless
  the vector field acquires a non-trivial profile, depending on specific integration constants. We commented on a physical
  interpretation of these results in terms of symmetries of the original action,
  and how the the vector profile can be probed by directly coupling matter and vectors, as for example in the
  interior of neutron stars. We also shown that  when including standard
  vector kinetic terms in the form of a Maxwell Lagrangian, 
   solutions for vector-Galileons in vacuum correspond to small modifications of Reissner-Nordst\"om configurations. 
  
  \smallskip
  
    These results are  important  
    to eventually describe gravitationally bound configurations  
in modified theories of gravity, such as black holes and  neutron stars,   including   realistic matter fields forming or surrounding the object. Future developments can occur in two directions. 
First,  it is important to classify in full generality
all consistent vector-tensor theories which propagate only four degrees of freedom, even in absence of 
  Abelian gauge symmetry. These more general theories, interesting for phenomenology, can be equipped by 
   some alternative gauge symmetries,  or simply enjoy second class constraints which
    prevent the propagation of additional modes. The lessons learned in studying these classes
     of theories can then be used for other contexts, as in scalar-tensor  systems or, optimistically, to find explicit examples
     of partially massless massive gravity models. 
Second, our findings can be useful for 
 further studying the physics of gravitationally bound compact objects in modified gravity, and for relating their properties
 to symmetries or features 
  of the  systems one examines. The existence of stealth solutions ensures that the starting background configuration coincides with the one of Einstein gravity, but can lead to sizeable  differences   when considering the dynamics of fluctuations, or
  when coupling vector and matter fields. We hope to report soon on new results  along these directions.

\subsection*{Acknowledgments}

It is a pleasure to thank Marco Crisostomi, Carlos N\'u\~nez, and Ivonne Zavala for suggestions. JC is supported by CONACyt grants 263819 and 179208.

\begin{appendix}
\section{Disformal transformation and symmetries}\label{app-A}

In this Appendix, we show that  the vector-tensor Lagrangians 
 \bea
L_2&=& m^3\,\sqrt{-g}\,\sqrt{X} \,,\label{L2app}
\\
\tilde L_4&=& m\,\sqrt{-g}\,\left\{
 \sqrt{X} R +  \frac{1}{2 \,\sqrt{X}}\left[(\nabla_\mu A^\mu)^2 - \nabla_\rho A_\sigma \nabla^\sigma A^\rho   -
 \frac14\,F_{\mu\nu}
 F^{\mu\nu}
 \right]
\right\}\,,\label{L4app}
\eea
can be obtained from a certain limit of a disformal transformation acting on an Einstein-Hilbert system. This is  useful
to exhibit a new symmetry that these Lagrangians satisfy. 

\subsubsection*{Details of the disformal transformation}

We examine the following disformal transformation of the metric \cite{Bekenstein:1992pj}, which involves vector degrees of freedom \cite{Kimura:2016rzw} (but see also \cite{Bettoni:2013diz} for important papers discussing consequences of disformal transformations in scalar-tensor systems):
\be \label{bargmn}
g_{\mu\nu}\,\,\to\,\,\bar g_{\mu\nu}\,=\,g_{\mu\nu}- \frac{1}{\epsilon^2\,m^2}\, A_{\mu}A_{\nu}\,,
\ee
with inverse
\be
 \bar g^{\mu\nu}
\,=\,g^{\mu\nu}+\frac{\gamma_0^2}{m^2}\,A^{\mu}A^{\nu}\,.
 \ee
 We define
   \be
 \gamma_0^2\,=\,\frac{m^2}{m^2\,\epsilon^2+2 \,X}\,,
 \ee
 and $2 X=-A^2$. 
 Here $m$ is a mass scale (for definiteness, the same mass scale appearing in eqs \eqref{L2app}, \eqref{L4app}), and $\epsilon$ an arbitrary dimensionless  parameter, which  we 
  consider as  small for the sake of our arguments.
  We apply the disformal transformation to 
a Lagrangian made of Einstein-Hilbert, Maxwell, and cosmological constant terms:
\bea
\epsilon\,{ L}&=& \sqrt{-g} \left[ \epsilon \,M_{Pl}^2\,R+ \,\epsilon\,\Lambda-\frac{1}{4\,\epsilon}\,F_{\mu\nu} F^{\mu\nu}\right]\,.
\eea
 We weight parts of the Lagrangian density with powers of the parameter $\epsilon$ before performing the transformation,
 this will be important in what follows.

The disformal transformation, when applied to the cosmological constant term, gives
 \bea \label{defgdis}
  \epsilon\,\Lambda\, \sqrt{- g}\,\to\,
\epsilon \,\Lambda\, \sqrt{-\bar g}&=& \gamma_0^{-1}\,\Lambda\, \sqrt{-g}
\\
&=& \frac{\,\Lambda\,}{m}\,{\sqrt{2 X}}\,\sqrt{-g}+{\cal O}(\epsilon^2) \,. \label{L2disf}
 \eea
 In the $\epsilon\to0$ limit, the resulting Lagrangian after the transformation is
 $L_{(2)}$, given in eq \eqref{L2app}.

 \smallskip
 
While for the Einstein-Hilbert plus Maxwell Lagrangian, we find that the disformed contribution is 
\be \label{defRdis}
{\epsilon}\,M_{Pl}^2\,
\sqrt{- g} {R}-\frac{\sqrt{-g}}{4 \epsilon}\,F_{\mu\nu} F^{\mu\nu}\,\to\,
{\epsilon}\,M_{Pl}^2\,
\sqrt{-\bar g} \bar{R} - \frac{\sqrt{-\bar g}}{4 \epsilon}\,F_{\mu\nu}  F_{\rho\sigma} \bar{g}^{\mu \rho} \bar{g}^{\nu \sigma}\,,
\ee
with 
\begin{align}
\sqrt{-\bar g}\left[{\epsilon}\,M_{Pl}^2\,
 \bar{R}  \right.&\left.- \frac{1}{4 \epsilon}\,F_{\mu\nu}  F_{\rho\sigma} \bar{g}^{\mu \rho} \bar{g}^{\nu \sigma}\right] = 
  \sqrt{-g} \,\frac{M_{Pl}^2}{m} \left\{ \,\gamma_0^{-1}\,R + \gamma_0\, \left[
(\nabla_\mu A^\mu)^2-\nabla_\mu A^\nu \nabla_\nu A^\mu-\frac14 F^2
\right]
 \right\}
\\=&  \sqrt{-g} \,\frac{M_{Pl}^2}{m} \left\{ \,\sqrt{2 X}\,R + \frac{1}{\sqrt{2 X}}\, \left[
(\nabla_\mu A^\mu)^2-\nabla_\mu A^\nu \nabla_\nu A^\mu-\frac14 F^2
\right]
 \right\}+{\cal O}(\epsilon^2) \,, \label{L4disf}
\end{align}
%
plus total derivatives. In the previous expression, $F^2=F_{\mu\nu}F^{\mu\nu}$. 
%
 In the $\epsilon\to0$ limit, the resulting Lagrangian is
 $\tilde L_{(4)}$ given in eq \eqref{L4app}.

\subsubsection*{An heuristic derivation of a new  symmetry}

Hence the initial Einstein-Maxwell action, equipped with a cosmological constant, is disformally equivalent to a set of
vector-tensor theories, labelled by a dimensionless quantity $\epsilon$ which parameterises the disformal transformation. 
 The original action has an Abelian gauge symmetry, while the final system  is not Abelian
 symmetric. Nevertheless, we would expect that some form of memory of the original symmetry remains. Indeed, 
  we are going to show that a  new gauge symmetry arises for the disformed theory. 

The disformal transformation we are examining is built in terms of the tensor object
\be\label{bargmn2}
g_{\mu\nu}\,\,\to\,\,\bar g_{\mu\nu}\,=\,g_{\mu\nu}- \frac{1}{\epsilon^2\,m^2}\, A_{\mu}A_{\nu}\,.
\ee
The vector does not  necessarily transform under disformal transformation. On the other hand, 
recalling that the original theory is invariant under gauge transformation, being built in terms  of $F_{\mu\nu}$, we can disformally map the vector to    
\be\label{bargmn3}
A_\mu\to \bar A_\mu = A_\mu+\partial_\mu f
\,,
\ee
for an arbitrary function $f$: 
the result of the disformal transformation is independent from $f$.
%
 

These two facts  imply that
{\it any}  transformation acting on the fields $g_{\mu\nu}$, $A_\mu$, which leaves invariant expressions \eqref{bargmn2},  \eqref{bargmn3},  is  a symmetry of the system, obtained
 after 
applying the disformal transformation.
An example of  symmetry is:
\bea \label{newsi1}
A_\mu&\to&A_\mu+\epsilon^2\,m^2\,\partial_\mu \xi\,,
\\
\label{newsi2}
g_{\mu\nu}&\to&g_{\mu\nu}+\partial_\mu \xi A_{\nu}+ A_{\mu}\,\partial_\nu \xi +\epsilon^2\,m^2\,\partial_\mu \xi \partial_\nu \xi\,,
\eea
for an arbitrary scalar field $\xi$ (with dimension inverse of the square of a mass). Notice that the vector transforms in eq \eqref{newsi1}
 as an Abelian gauge transformation. The limit $\epsilon\to0$ of this transformation is  well defined, and consists of a transformation which only acts on the metric:
\bea \label{symAp}
g_{\mu\nu}&\to&g_{\mu\nu}+ A_{\nu}\partial_\mu \xi + A_{\mu}\,\partial_\nu \xi\,.
\eea
Hence transformation \eqref{symAp} is a symmetry for the Lagrangians of eqs \eqref{L2app}, 
\eqref{L4app}. It would be interesting to develop this method, based on disformal transformations, to find new symmetris for more general scalar Horndeski and vector
tensor actions starting from specific Abelian symmetric actions.


\end{appendix}

\end{document}